\documentclass[a4paper,twocolumn,prl,longbibliography]{revtex4-1}
\usepackage[utf8]{inputenc}
\usepackage[english]{babel}
\usepackage{amsmath}
\usepackage{amsfonts}
\usepackage{amssymb}
\usepackage{graphicx}
\usepackage[left=2cm,right=2cm,top=2cm,bottom=2cm]{geometry}
\usepackage{tikz-feynman,contour}
\usepackage{slashed}
\newcommand{\cD}{\mathcal{D}}

\newcommand{\cM}{\mathcal{M}}
\newcommand{\cP}{\mathcal{P}}
\newcommand{\va}{\mathbf{a}}
\newcommand{\vb}{\mathbf{b}}
\newcommand{\vk}{\mathbf{k}}
\newcommand{\vR}{\mathbf{R}}

\newcommand{\vp}{\mathbf{p}}

\newcommand{\R}{\mathbb{R}}
\newcommand{\bra}{\langle}
\newcommand{\ket}{\rangle}
\begin{document}
\title{Poincar\'e group spin networks}
\author{Altaisky M.V.}
\affiliation{Space Research Institute RAS, Profsoyuznaya 84/32, Moscow, 117997, Russia}
\date{Revised: June 5, 2024}
\begin{abstract}
Spin network technique is usually 
generalized to relativistic case by changing $SO(4)$ group -- Euclidean counterpart of the Lorentz 
group  -- to its universal spin covering $SU(2)\times SU(2)$, or by using the representations of $SO(3,1)$ Lorentz group. 
We extend this approach by using {\em inhomogeneous} Lorentz group $\cP=SO(3,1)\rtimes \R^4$, which results in the simplification of the spin network technique. The labels on the network graph corresponding to the subgroup of translations $\R^4$ make the 
intertwiners into the products of $SU(2)$ parts and the energy-momentum conservation delta functions. This maps 
relativistic spin networks to usual Feynman diagrams for the matter fields.  

\end{abstract}
\maketitle
\section{Introduction} 
Having been the most fundamental problem of physics over a hundred years the problem of quantization of gravity -- construction 
of a quantum theory of spacetime, consistent with both the quantum field theory (QFT) and general relativity (GR) -- still remains 
unsolved. Both sides of this problem are related to certain symmetry groups. The GR is a theory invariant under the general 
coordinate transformations Diff(4). Modern QFT of elementary particles is a gauge theory invariant under $SU(3)\times SU(2) \times U(1)$ group of internal symmetries and the Poincare symmetry group of the 4d Minkowski spacetime. A promising way of 
unification of these two approaches seems to be the loop quantum gravity (LQG) \cite{Rovelli1991CQG,RS1994,Rovelli-book}.
The quantization of gravity program may be traced since Dirac and Arnowitt-Deser-Misner represented the Einstein-Hilbert action 
\begin{equation}
S_{EH}[g] = \frac{c^3}{16\pi G} \int d^4x \sqrt{|g|} R , \label{EH} 
\end{equation}
where $R$ is the Ricci scalar, $G$ is gravity constant, and $g \equiv  \mathrm{det}(g_{\mu\nu})$, in Hamiltonian form \cite{Dirac1958,ADM1959}, formally suitable for quantization.

Usual canonical quantization have not succeed in gravity theory for a number of reasons. First of all, preferred role 
the time coordinate $t$ plays in canonical quantization contradicts the general symmetry between space and time coordinates implied by GR. The second trouble, which may seem more philosophical than practical, is that the Einstein-Hilbert action 
\eqref{EH} describes ''geometry without matter,'' which does not exist since we \textcolor{red}{consider} space-time is but a set of relations 
between matter fields, and hence cannot exist without matter. 

Significant progress in quantization of gravity has been achieved when Ashtekar has represented Einstein-Hilbert 
action in a form of non-Abelian gauge theory, with the gauge field being the self-dual part of the spin connection  \cite{Ashtekar1986,Ashtekar1987}:
\begin{equation}
A^{IJ}_\mu[e] = w^{IJ}_\mu[e] - \frac{\imath}{2}\epsilon^{IJ}_{\phantom{IJ}MN} w^{MN}_\mu[e],
\end{equation}
where $w^{IJ}_\mu[e]dx^\mu$ is a spin connection, and $e$ stands for tetrads. Since the Ashtekar connection $A$ takes 
its values in the $su(2)$ Lie algebra, its geometric interpretation suggests that the quantum state of ''pure spacetime'' 
given by $A$ can be described by wave function $\Psi[A]$. This function can be expanded in a given basis in the Hilbert 
space of states. This basis may be a basis of loop states $|\alpha\ket$. Let 
\begin{equation}
U(A,\alpha) = P e^{\oint_\alpha A},\quad T[A,\alpha]:=\mathrm{Tr} U(A,\alpha)
\end{equation}
be the holonomy operator corresponding to the action of the connection $A$ along a loop $\alpha$, where $P$ means path ordering, then the basis of all possible loops $\{\alpha\}$ coordinatize the phase 
space of GR in a sense that any holomorphic and gauge invariant functional of $A$ can be expressed as functional 
of $T[A,\alpha]$ \cite{Jacobson1988,RS1990}.

However, the Ashtekar spin-connection formalism in a quantum theory does 
not provide the same level of consistency between the geometry ($g_{\mu\nu}(x)$) and the matter fields ($\psi(x)$) that we have in a classical theory.
In classical theory, the variation of the full action 
$S[g,\psi] = S_{EH}[g] + S_{Matter}[g,\psi]$ leads to the Einstein equations 
\begin{equation}
R_{\mu\nu} - \frac{1}{2}g_{\mu\nu} R = \frac{8\pi G}{c^4} T_{\mu\nu}.
\label{ee}
\end{equation}
The r.h.s. of the Einstein equations \eqref{ee} is proportional to 
the energy-momentum tensor of the matter fields. 
This means, at least in principle, the Riemann tensor $R_{\mu\nu}$ in the l.h.s. can be expressed in 
terms of the matter fields, and the spacetime it accounts for turns to be 
merely a set of relations between the matter fields.

In quantization frameworks, known to the author, the geometry of spacetime 
described by the Einstein-Hilbert action \eqref{EH} is treated independently of the matter fields. Its variation gives the ''empty 
space equations''
$$
R_{\mu\nu}-\frac{1}{2}g_{\mu\nu}R=0,
$$
and the matter fields are added to gravity model fixed by $g_{\mu\nu}$ 
(regardless whether or not the given model is considered as ''background-independent'').

The present paper makes an endeavour to construct quantum space-time geometry directly in terms of matter fields.

\section{Spin networks} 
Spin network techniques is intimately related to the Regge discrete gravity in three dimensions \cite{Regge1961}.
In asymptotic limit, the triangulation of a given 3d manifold $\cM$ can be used to approximate the Einstein-Hilbert action 
$\int_\cM d^3x \sqrt{g}R$ by the Regge action 
\begin{equation}
S_{R} = \sum_{i=1}^6 \theta_i \left(j_i + \frac{1}{2}\right), \label{SR}
\end{equation} 
where $j_i + \frac{1}{2} \gg 1$ is the length of the edge $j_i$ of the tetrahedron taken in certain appropriately 
chosen small units, $\theta_i$ is the angle between two outer normals to the surfaces joining at this edge. The full Einstein-Hilbert action is then approximated by the sum over all tetrahedra in the triangulation of $\cM$ 
\cite{Regge1961,WT1992}. For a given tetrahedron with a set of parameters $(j_1,\ldots, j_6)$ the Ponzano-Regge action \eqref{SR} can 
be asymptotically expressed as \cite{Ponzano1968}
\begin{equation}
(-1)^{\sum_{i=1}^6 j_i} \left\{
\begin{matrix}
j_1 & j_2 & j_3 \cr
j_4 & j_5 & j_6
\end{matrix}
\right\} \sim \frac{1}{\sqrt{12\pi V}} \cos \left( S_{R}+\frac{\pi}{4}\right),
\end{equation}
where $V$ is the volume of the given tetrahedron, and $\left\{\begin{matrix}
j_1 & j_2 & j_3 \cr
j_4 & j_5 & j_6
\end{matrix}\right\}$ are the Racah-Wigner 6j symbols for the angular momentum \cite{Brink}.  In four dimensions,  
triangulation is performed with 4d simplices and the discrete action can be expressed in terms of 10j or even 15j symbols 
\cite{CKY1997}.

For an arbitrary curved manifold with the triangulation graph $\Gamma$ the discretization of geometry can be done by assigning 
the indices of a group representation $j_i$ to each edge of $\Gamma$. According to the Penrose toy model, 3d 
geometry emerges in a system of spin-$\frac{n\hbar}{2}$ particles if we consider a graph  in which the matter particles 
are considered as edges and the vertices are constructed in a combinatorial way so that the angular momentum is conserved 
in each vertex \cite{Penrose1971a}. 
It is possible to determine angles in the Penrose's 3d geometry, but not the lengths, except for a use of angular momentum 
to determine the distance between particles in a composite object \cite{Szabados2022}. 
Technically, the Penrose's idea can be extended to an arbitrary triangulation graph $\Gamma$, 
with the edges corresponding to the action of the connection $A$ of the group $G$ taken in $j_i$ representation, and the 
vertices (intertwiners) are constructed to make singlets of the edges being connected \cite{RS1995}. This is specially suitable for 
$SU(2)$ group, for which all representations are constructed by appropriate symmetrization of fundamental representations. 
If $\Gamma$ is such graph, and $\{\gamma_i,\}_{i=\overline{1,L}}$ is the set of its edges, the holonomy operator, 
or the parallel propagator along the edge $\gamma_i$, 
\begin{equation}
U(A,\gamma_i) := P e^{\int_{\gamma_i}A} \label{gi}
\end{equation}
defines a space of functions over the graph
\begin{equation}
\Psi_{\Gamma,f}[A] := f(U(A,\gamma_1),\ldots,U(A,\gamma_L)),
\end{equation}
where $f(\cdot,\cdots,\cdot)$ is a smooth function of all its arguments. The scalar product in the space of functions over 
graphs is defined by integration over the group $G$ with an appropriate Haar measure:
\begin{equation}
\bra \Psi_{\Gamma,f}|\Psi_{\Gamma,g}\ket\!=\!\int dU_1\!\ldots\!dU_L 
\overline{f(U_1,\ldots,U_L)}g(U_1,\ldots,U_L) \label{scu}
\end{equation}
The triplet $S=(\Gamma,j_i,i_n)$ is called a spin network (SN) embedded in continuous manifold $\cM$. For a closed 
graph $\Gamma$ the scalar product \eqref{scu} [at $f\!=\!g$] defines a norm of spacetime configuration given by $\Gamma$, 
i.e. the probability that such spacetime can emerge. A norm of an open graph is obtained by closing two copies of 
the spin network to a closed graph $\Gamma\#\tilde{\Gamma}$ \cite{Penrose1971a,BC1998}.

The ''quantumness'' of the Penrose 3d toy model consists in the observation, that classical geometry -- angles between two directions -- emerges in large $N$ limit in the gedanken experiment in which a single spin-half fermion is voluntary transferred 
from one large $N$-block to another $M$-block. Time does not play any explicit role in such consideration and its generalization to four dimensions usually done in the framework of {\em spin foam} is not 
straightforward, see e.g. \cite{Baez1998,Oriti2001} for a discussion. It is a point to note that the Einstein-Hilbert action \eqref{EH} appears meaningless without matter fields and 
that spacetime should be considered as a set of states of matter fields, corresponding to physical events. In this respect the 
SN seem more fundamental than the classical action \eqref{EH}. Indeed, considering each edge $j_i$ of 4d SN as a {\em physical particle} and the intertwiners $i_k$ as functionals describing the changes occurring to particles at physical events, we necessarily  have to fulfil the correspondence principle: The newly constructed quantum theory should turn to the known 
quantum field theory, invariant under special relativity transformations when the gravity constant vanishes. 

The basic idea beneath the generalization of the Penrose's spin networks to four dimensions is the change of tetrahedra of the 3d 
Regge gravity into 4d simplices of four-dimensional model followed by the replacement of $SO(3)$ rotation group by $SO(4)$, or 
by its universal covering $SU(2)\times SU(2)$. This is called {\em relativistic spin network} \cite{BC1998}. The product 
$SU(2)\times SU(2)$ labels triangles, i.e. {\em bivectors}, rather than the edges. Alternatively, the group $SL(2,\mathbb{C})$, 
which is the covering group of the Lorentz group, can be considered. This corresponds to writing a general element of the 
Lorentz algebra as a sum of a rotation $J$ and a boost $K$. Then $J+\imath K$ is an element of the algebra of complex 
rotations $so(3,\mathbb{C})$ \cite{BC2000}. In both cases the Lorentz rotations are associated with bivectors in the triangulation of 
4-manifold:
$$
b^{ab}=\frac{1}{2}\epsilon^{abcd}L^e_{\phantom{e}d}g_{ec}.$$  
This is in agreement with Rovelli-Smolin spin networks, where fundamental operators act on faces, rather than edges \cite{RS1995}.

The starting point of either of such definitions of relativistic spin networks is the Lorentz invariance of the Regge-Ponzano action defined on a 4-manifold. We can make this starting point more concrete by making an assertion that quantum spacetime, 
being a collection of all possible physical events, should be invariant under the Poincare group $\cP$, rather than merely a homogeneous Lorentz group $SO(3,1)$. 
Indeed, the identification of particles with the representations of the inhomogeneous Lorentz group has a long history, 
starting from Wigner \cite{Wigner1939} in the context of special relativity. Roughly speaking, particles are identified with a 
set of observable properties, subjected to symmetry transformations when different observers, dwelling in different frames of reference talk, to each other, with the probabilities of physical processes remaining the same for either of the observers 
\cite{Weinberg1}.  
This means, the laws of physics, tested locally, should be the same for all laboratories, 
regardless their positions and their velocities.

In classical special relativity, each observer is associated with a Lorentzian frame of reference and the coordinates are 
transformed between observers according to the Lorentz transformation law. In quantum case we ought to assume there 
should be no frame of reference without matter, and thus associate any frame of reference with a matter particle, labelled 
by a representation of inhomogeneous Lorentz group. 

Starting from the general definition \eqref{gi} this implies that operators $A$, describing transformation of matter fields, 
should be taken in appropriate representation of the Poincare group $\cP$. Since each edge of 4d triangulation graph $\Gamma$ 
is identified with a matter particle, the representation of the group $\cP$ should be multiplied by appropriate representation 
of the internal symmetry group (say, $SU(3)\times SU(2)\times U(1)$, but this will be dropped for bookkeeping purposes).
From this point we do not need to include the ''pure space'' action \eqref{EH} by hand any longer.  

\section{Representations of Poincar\'e group} 
Poincare group is a semidirect product of the Lorentz group and translation group. 
The action of the Poincare group $\cP$ on arbitrary vector $\hat{e}$ in Minkowski space consists of a 
Lorentz transform $L$ followed by a translation  on a constant vector $\hat{\varepsilon}$:
\begin{align}
P(\hat{\varepsilon},L) \hat{e} &= L \hat{e} + \hat{\varepsilon}, \quad \hat{e},\hat{\varepsilon} \in \mathbb{R}^{3,1}, \\
\nonumber P(\hat{\varepsilon}_2,L_2)P(\hat{\varepsilon}_1,L_1)&= P(\hat{\varepsilon}_2+L_2\hat{\varepsilon}_1,L_2L_1)
\end{align} 
The subgroup of translations $\R^4$ is abelian group. Its representations are one-dimensional and can be labelled 
by vectors of the Minkowski space. A vector $|k\ket$ in a Hilbert space of states transforms under spacetime translations 
as 
\begin{equation}
T(\hat{\varepsilon}) |k\ket = e^{\imath \hat{k}\cdot\hat{\varepsilon}} |k\ket.
\end{equation}
All vectors $k'$, which can be obtained from a fixed vector $k$ by Lorentz rotation
$\hat{k}'=L\hat{k}, \quad |\hat{k}'\ket = T(L)|\hat{k}\ket,$
transform under spacetime translations in the same way:
$$
T(\hat{\varepsilon}) T(L) |\hat{k}\ket = e^{\imath (L\hat{k})\cdot \hat{\varepsilon}} T(L)|\hat{k}\ket.$$

The simplest case corresponds to $\hat{k}=0$. In this case all vectors in Hilbert space are invariant with respect 
to translations, and the representations of the Poincare group are effectively reduced to the representations $L^{(j,j')}$ of 
the Lorentz group \cite{GMS1963}. This case corresponds to massless particles.

The next case corresponds to representations determined by a time-like vector 
\begin{equation}
\hat{k}_0 = (k_0,0,0,0), \label{k0}
\end{equation}
describing a particle of rest mass $m_0=\frac{\hbar k_0}{c}$. Since vector \eqref{k0} is invariant 
under spatial rotations, the Hilbert space vectors of the form $T(\mathrm{R}(\mathbf{a}))|\hat{k}_0\ket$ will transform 
as $e^{\imath \hat{k}_0\cdot\hat{\varepsilon}}$ under translations. The vectors obtained by all spatial rotations of $\hat{k}_0$ form invariant subspace and can be labelled by spin $s$ and its projection onto a given axis $m_s$. Under spatial rotations such 
vectors transform according to corresponding representation of the rotation group:
\begin{equation}
T(\mathrm{R}(\va))|\hat{k}_0 s m_s\ket = \sum_{m_s'}D^{(s)}_{m_s' m_s}(\va)|\hat{k}_0 s m_s'\ket.
\end{equation} 
For a fixed value of $\hat{k}_0$ and fixed value of spin $s$ there 
are $2s+1$ vectors forming a subspace invariant under rotations. Using all possible Lorentz transformations of $\hat{k}_0$ 
we can form a representation of the full Poincare group:
\begin{equation}
|\hat{k}s m_s\ket = T(Q[\vb(\vk)]) |\hat{k}_0 s m_s\ket,
\end{equation}
where $\hat{k} = L \hat{k}_0$. The ''length'' is preserved by Lorentz transform $\hat{k}\cdot\hat{k}= \hat{k}_0\cdot \hat{k}_0=k_0^2$. 
This means, each representation of the Poincare group is labeled by rest mass $k_0$ and spin $s$ \cite{ED1979}:
\begin{align}
T(\varepsilon,L)  |\hat{k}s m_s\ket = e^{\imath \hat{k}'\cdot \hat{\varepsilon}} \sum_{m_s'} D^{(s)}_{m_s' m_s}(\vR') 
|\hat{k}' s m_s\ket,\\
 \nonumber 
\hat{k}'=L\hat{k} = Q'\hat{k}_0, 
\end{align} 
Any vector in Hilbert space of states can be expanded in the bases of such representations by summing up over $m_s$ and integrating 
over all possible Lorentz boosts, given by different vectors $k$ with the same norm (rest mass):
\begin{equation}
|\psi\ket = \sum_{m_s} \int \psi_{m_s}(\hat{k}) |\hat{k} s m_s\ket \frac{d^3\vk}{2k_t},\quad \hat{k}\equiv(k_t,\vk).
\label{lm}
\end{equation}
For a given value of rest mass ($k_0$) these basic vectors form an orthogonal basis:
\begin{equation}
\bra \hat{k}' s m_s'|\hat{k} s m_s\ket = 2k_t \delta_{m_s m_s'} \delta^3(\vk-\vk').
\end{equation}

If we assume that spacetime is but a set of relations between the {\em matter} fields, we have to avoid the 
assumption of ''pure space'', associated with the Einstein-Hilbert action \eqref{EH}, and we ought to substitute the 
representations of $SU(2)$ group on the edges of the original Penrose's spin network by appropriate 
representations of  matter fields symmetry group, {\sl viz.} Poincare symmetry group $\cP$ times the group of 
internal symmetries $G$. The spacetime is then a discrete collection of particles -- edges of the graph $\Gamma$, labelled 
by the indices of appropriate representations of Poincare group $\cP$ and the indices of internal symmetry group $G$. 
These particles are joined together by three-valent vertices, the intertwiners -- spin-tensors constructed so that 
their convolution with all joining edges form singlets with respect to both $\cP$ and $G$. The graph $\Gamma$ with edges 
labelled by these representation defines the state of spacetime in a Hilbert space of states with the scalar product 
\eqref{scu}, $U \in \cP\times G$. The norm of the state $\bra \Gamma|\Gamma\ket$ is calculated in a usual way 
by joining the open ends of graph $\Gamma$ with conjugated graph and integrating over all edges \cite{Penrose1971a}.

The representations of Lorentz group for particles with non-zero mass are given by spin tensors. In the simplest 
case of spin-$\frac{1}{2}$ particles this is represented by left- and right-spinors 
\begin{align*}
\psi_A \to \psi_A' = e^{\frac{\imath}{2}(\mathbf{w}-\imath \boldsymbol{\nu})\cdot \hat{\sigma}} \psi_A, \\
\psi_B \to \psi_B' = e^{\frac{\imath}{2}(\mathbf{w}+\imath \boldsymbol{\nu})\cdot \hat{\sigma}} \psi_B,
\end{align*}
where $\hat{\sigma}$ are the Pauli matrices.
The invariant measure on the group of $SU(2)$ rotations $U_\mathbf{w}(\psi_w)=e^{\imath\frac{\psi_w}{2}\mathbf{w}\cdot \boldsymbol{\sigma}}$ is 
\begin{equation}
d\mu ( U_\mathbf{w}(\psi_w) ) = \frac{1}{2} \sin^2 \left(\frac{\psi_w}{2} \right) d\psi_w \sin \theta_w d \theta_w d\phi_w,
\end{equation}  
where $(\theta_w,\phi_w)$ are the Euler angles, determining the direction of vector $\mathbf{w}$, and $\psi_w$ is the rotation angle.

If we assume that the properties of spacetime are completely determined by symmetry properties 
of matter particles, we should consider a graph $\Gamma$, the edges of which are labelled 
by representations of Poincare group $\cP$, and the vertexes being the intertwiners making singlets of all joining edges by convolution in appropriate arguments. The norm of each graph $\Gamma$, or equivalently of each possible scenario of the 
spacetime, is obtained by integration in \eqref{scu} over all edges $\prod_{i=1}^E d\mu \left(\cP^{(i)} \times G^{(i)} \right)$, 
where $\cP^{(i)}$ and $G^{(i)}$ are representations of the Poincare group and the internal symmetry group on the 
$i$-th edge, with $f=g$ determined by the set of vertices of $\Gamma$. 

The measure on the Poincare group can be constructed as a product of the measures on the proper orthochronous Lorentz 
group $SO(3,1) \cong SL(2,\mathbb{C})/\mathbb{Z}_2$ and the measure on the translation group $\R^4$.
The group $SL(2,\mathbb{C})$ can be represented as a direct product \cite{Takahashi1963,AV1999}: 
$$
SL(2,\mathbb{C})=SU(2)\times \mathbb{R}^+ \times \mathbb{C}.
$$
Each matrix $\|a_{ij}\|\in SL(2,\mathbb{C})$ can be casted 
in a form 
\begin{equation}
\begin{pmatrix}
a_{11} & a_{12} \cr a_{21} & a_{22}
\end{pmatrix} = 
\begin{pmatrix}
u_{11} & u_{12} \cr u_{21} & u_{22}
\end{pmatrix} \times 
\begin{pmatrix}
\lambda^{-1} & \mu \cr 0 & \lambda
\end{pmatrix}, \label{sl2f}
\end{equation}
where $\|u_{ij}\|\in SU(2), \mu \in \mathbb{C}, \lambda>0$. 
The triangular matrices in the r.h.s. of Eq.\eqref{sl2f} form a normal subgroup in $SL(2,\mathbb{C})$, which determines 
Lorentz boost. The complex number $\mu$ determines the direction of boost by means of the projection of complex plane 
onto the sphere $S^2$ \cite{AV1999}, and $\lambda$ determines the hyperbolic angle of the boost. 
The parameters of the boost can be expresses in terms of a 4d vector  ($p_0^2 - \vp^2 = M^2$), such that :
\begin{align*}
p_0 &= M \cosh \eta,\\
p_3 &= M \sinh \eta \cos \theta, \\
p_1 &= M \sinh \eta \sin \theta \cos \phi, \\
p_2 &= M \sinh \eta \sin \theta \sin \phi.
\end{align*}
The boost parameters then become 
\begin{equation}
\lambda = \left( \frac{p_0-p_3}{2M} \right)^\frac{1}{2},
\mu = \frac{2}{M\lambda}(p_1 + \imath p_2) \equiv \mu_1+\imath \mu_2, 
\end{equation}
The measure on the group of triangular matrices $\begin{pmatrix}
\lambda^{-1} & \mu \cr 0 & \lambda
\end{pmatrix}$, that is 
\begin{equation}
d\mu_L = \frac{\lambda d\lambda}{4} d\mu_1 d\mu_2,
\end{equation}
then becomes 
\begin{equation}
d\mu_L = \frac{1}{2} \sinh^2 \eta d\eta \sin\theta d\theta d\phi.
\end{equation}
The integration over all boosts (within a representation given by $M$) is 
\begin{align*}\nonumber 
d^4p \delta(\hat{p}^2-M^2) = \frac{d^3p}{2\sqrt{\vp^2+M^2}}= \frac{p^2dp \sin\theta d\theta d\phi}{2\sqrt{\vp^2+M^2}}\\
=\frac{M^2}{2}\sinh^2 \eta d\eta \sin\theta d\theta d\phi.
\end{align*}
The latter equation coincides with the usual integration measure on the lines of Feynman diagrams. 

If we include the representation of the translation group $\R^4$ on each edge, $e^{\imath \hat{p}\hat{y}}$, and integrate over translation 
coordinates $\hat{y}$, we yield the factor $\delta(\sum_i \hat{p}_i)$ in each vertex, and thus the norm of the 
graph $\Gamma$, calculated according to the rule \eqref{scu} of spin networks coincides with the value of the Feynman diagram 
corresponding to this graph. 

\section{Feynman diagrams}
The relations between spin foam models and Feynman diagrams have been already considered in \cite{BF2007,Baratin_2007,Freidel2019}.
In special relativity the observable quantities -- amplitudes of physical processes modulus squared -- are 
related to Feynman diagrams 
\begin{equation}
I_\Gamma = \int_{\R^4} d^4x_1\ldots d^4x_n \prod_{(ij)\in\Gamma} G_F(|\vec{x}_i-\vec{x}_j|), \label{A}
\end{equation}
where $G_F(|\vec{x}_i-\vec{x}_j|)$ is the Feynman propagator along the given edge of the graph $\Gamma$. Each integration 
variable $x_i$ corresponds to a vertex of the graph $\Gamma$.
The integration over all 4d Minkowskian or Euclidean space is implied.

According to the classical correspondence rules between the special relativity and the general relativity, 
the transition to GR is performed by changing the flat space $\R^{3,1}$ into a curved manifold supplied with a 
given metric $g_{\mu\nu}(x)$. This manifold may be triangulated by appropriate graphs with each edge being 
ascribed by a length $l_{ij}$ in accordance to the metric $g$. The usual object of interest is 
\begin{equation}
\tilde{I}_\Gamma(l_P) = \int \cD g e^{\frac{\imath}{l_P}S(g)} I_\Gamma(g), \label{B}
\end{equation} 
that is the same Feynman diagram \eqref{A}, but with the edge lengths given by $g$. weighted by a 
''pure gravity'' action $S(g)$, and integrated over all possible metrics.

We know that a continuous differentiable manifold with a metric $g$ is merely an approximation of 
a discrete set of events, linked to each other by Feynman's propagators. If a curved manifold is embedded into 
a flat space  it may be triangulated, so that each triangle is flat, and all curvature is concentrated in vertices.
A 3d compact manifold can embedded in $\R^4$. Its triangulation is made of tetrahedra (3d simplices). Similarly, a 4d 
compact manifold can be triangulated by 4d simplices. 
The integration measure on an elementary graph \eqref{A} ($n=4$), corresponding to a single tetrahedron, can 
be casted in a form
\begin{equation}
d^4x_1\ldots d^4x_4 =  d\Lambda d^4a \prod_{i<j} l_{ij} dl_{ij},
\end{equation} 
where $d\Lambda d^4a$ is the Haar measure on the inhomogeneous Euclidean $SO(4)\rtimes\R^4$ (resp., Lorentz $SO(3,1)\rtimes \R^4$) group,
$l_{ij}$ are the lengths of the edges of the tetrahedron.
For a 4d triangulation containing more than 4 vertices, the integration measure can be written in a similar form 
$$
d^4x_1\ldots d^4x_{4+k} =  d\Lambda d^4a \sum_{\epsilon \in (\pm1)^k} \prod_{e\in \Delta_k} l_{e} dl_{e} 
\prod_{\sigma\in\Delta_k} \frac{1}{V_\sigma},
$$
where $\epsilon$ is the orientation of 4d simplex in the triangulation $\Delta_k$, $V_\sigma$ is the volume of 
a given simplex \cite{BF2007,Baratin_2007}.
The Haar measure $d\Lambda d^4a$ is globally factorised here due to the global invariance of the theory with respect to the Euclidean (or, resp.,  Poincare) group. The value of each Feynman diagram is therefore defined by the set of edge lengths of its graph $\Gamma$ in a given triangulation. 

The importance of the Poincare group, rather than its Euclidean counterpart $ISO(4)$, in loop quantum gravity basis 
have been stressed in \cite{Freidel2019}. That paper starts from the existence 
of a continuous theory 
invariant under diffeomporphisms and determined by the charges on its boundary, considered as quantum punctures.
In hydrodynamical language \cite{Freidel2019} implements an interesting, but rather special case when the vortex 
charges are conserved on the boundary surface. 
Combinatorial approach, which we follow in this paper, is in some sense opposite to it: we assume the existence of 
a set of quantum particles, described by representations of the Poincare group, and the continuous geometry should be 
derived from this set in the limit of large number of particles. 
In a combinatorial approach, which follows the lines of Penrose's ideas, the Poincare group measure $d\Lambda_e d^4a_e$ 
should be ascribed to each edge of the graph separately, rather than globally, as was done in \cite{Freidel2019}. This means, each material particle, which 
labels an edge of graph $\Gamma$, is associated with a potential frame of reference. The observations in such frames  
of reference are invariant under certain symmetry transformations. 

According to the definition \eqref{gi},  connection $A$ defines a parallel propagators $U(A,\gamma_i)$ along the edge $\gamma_i$.
For the Poincare group $\cP$ the propagators are known. In momentum representation we have 
\begin{equation}
G_F(k) = \imath\frac{\slashed{k}+M}{k^2-M^2},\quad G_B(k) = -\imath\frac{g_{\mu\nu} - \frac{k_\mu k_\nu}{M^2}}{k^2-M^2}
\end{equation}
for the propagators of a spin-half fermion of mass $M$, and a massive vector boson, respectively.

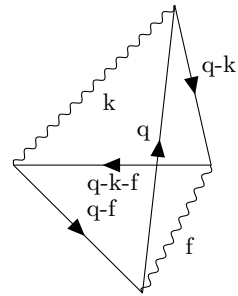
\begin{figure}[ht] 
\begin{tikzpicture}[baseline=(d)]
\begin{feynman}[horizontal = (a) to (b)]
\vertex (d);
\vertex [right =2.6cm of d] (a) ;
\vertex [below right =2.4cm of d] (b) ;
\vertex [above right = 3cm of d] (g) ;
\diagram*{
(g) -- [photon, edge label=k] (d) -- [fermion, edge label=q-f] (b),
(a) -- [fermion, edge label=q-k-f] (d), 
(a) -- [photon, edge label=f]  (b),
(g) -- [fermion, edge label=q-k] (a),
(b) -- [fermion, edge label=q] (g)
};
\end{feynman}
\end{tikzpicture}
\caption{Toy-model spacetime consisting of 4 fermions and 2 bosons}
\label{t3:pic}
\end{figure}

Figure~\ref{t3:pic} schematically depicts a spacetime consisting of 4 fermions and 2 bosons, which are 
related to each other by 4 intertwiners.  
The ($s\!=\!1$) boson in such representation is equivalent to a pair of spin-half fermion lines, as in usual 
Penrose's spin networks \cite{Penrose1971a}, see Fig.~\ref{fbv:pic}:
\begin{figure}[ht]
\centering\includegraphics[width=4cm]{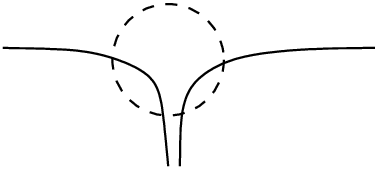}
\caption{Interaction of spin-$\frac{1}{2}$ fermions with spin-$1$ boson in spin network representation}
\label{fbv:pic}
\end{figure}
This implies the causal structure of quantum spacetime, say that shown in Fig.~\ref{t3:pic}, can be expressed by the combinatorics 
of fields represented by the edges of spin network. 
In this sequence of particles the bosons are multiply accounted 
according to the number of fermion lines they consist of. For the spin 1 massless photon this is an evident property, 
since for the photon propagator we have 
$$
\frac{1}{\slashed{k}}\frac{1}{\slashed{k}} = \frac{1}{k^2}.$$   
As in usual spin network formalism \cite{RS1995}, the quantum geometry can be casted in loop basis 
$\psi(\alpha)=\bra \alpha | \psi \ket$, where $\alpha$ is a loop, i.e., a sequence of particles labelled by their 
group representations. This forms a basis for a diffeomorphism-invariant theory required for quantization of 
general relativity \cite{RS1990}. 

Our consideration is, in some sense, similar to the {\em Poincare charge networks} described in \cite{Freidel2019}, with the difference 
that no string-like models are imposed, and the charges are not restricted to be on the boundary hypersurface. Instead, 
arbitrary combinations of particles are allowed -- with the Poincare group coordinates describing each particle in 
the network. 
In this respect the Poincar\'e group spin network combinatorics of particles is a discrete counterpart of 
general coordinate transformations of general relativity. Similarly to quantization of Arnowitt-Deser-Misner continuous model 
of gravity, we can expect constraints on vacuum state alike Wheeler-DeWitt equation \cite{DeWitt1967},
but such constraints should involve whole spin network rather than a three-dimensional subset.

\section{Conclusion}
In general relativity, the energy-momentum tensor in the r.h.s. of the Einstein equations enables one, 
at least in principle, to reconstruct the curvature of the spacetime. The solution will be dependent on the masses and 
velocities of the matter fields in the r.h.s. We have shown in this paper, that similar approach can be applied 
to quantum gravity, if we represent the quantum space-time by a spin network graph the edges of which are labelled by 
the representations of the Poincare group. Spin connection of such spacetime turns to be the function 
of masses and spins which label representations of the Poincare group on the edges of the spin network graph.

Simple discrete models, like that in Fig.~\ref{t3:pic}, can explain how the curvature depends on the masses, which label 
the edges of the spacetime graph. Taken alone, 
this cannot explain dynamics of the spacetime, which is related to the inflation of the Universe and consists in 
changing of the number of degrees of freedom in the network. This problem is postponed 
to forthcoming research. Possibly, it can be treated with known methods of the spin foam growth,
$$ 
\begin{tikzpicture}[baseline=(d)]
\begin{feynman}
\vertex (d);
\vertex [above =0.6cm of d] (c) ;
\vertex [below left = 0.7cm of d] (a) ;
\vertex [below right = 0.7cm of d] (b) ;
\diagram*{
(a) -- (d), (b)--(d), (c)--(d)
};
\end{feynman}
\end{tikzpicture}
\to 
\begin{tikzpicture}[baseline=(d)]
\begin{feynman}
\vertex (d);
\vertex [above =0.6cm of d] (c) ;
\vertex [above =0.3cm of d] (c1) ;
\vertex [below left = 0.8cm of d] (a) ;
\vertex [below left = 0.4cm of d] (a1) ;
\vertex [below right = 0.8cm of d] (b) ;
\vertex [below right = 0.4cm of d] (b1) ;
\diagram*{
(a) -- (a1) --(c1) --(c), (a1)--(b1) --(b), (b1)--(c1)
};
\end{feynman}
\end{tikzpicture}
$$
as described e.g. by Reisenberger and Rovelli \cite{Reisenberger1997}. 
 
\section*{Acknowledgement} The author is thankful to Prof. Yang-Hui He for useful references,  to Prof. S.Mikhailov for 
useful discussions, and to anonymous referee for useful comments. 
%

\end{document}